\documentclass[journal,transmag]{IEEEtran}
\usepackage{cite}
\usepackage{graphicx}
\usepackage{times}
\usepackage{latexsym}

\usepackage[mathscr]{eucal}
\usepackage{array}
\usepackage{fancyhdr}
\usepackage{amsmath,amssymb}
\usepackage{bm} 
\usepackage{citesort}
\usepackage{multirow}

\ifCLASSINFOpdf
\else
\fi

\newcommand{\vect}[1]{{\lowercase{\mathbf{#1}}}}

\newcommand{\m}{\vect{m}}

\usepackage{xcolor}

\makeatother

\usepackage{cite}
\usepackage{amsmath,amssymb,amsfonts}
\usepackage{algorithmic}
\usepackage{graphicx}
\usepackage{textcomp}
\usepackage{xcolor}
\usepackage{booktabs}
\usepackage{xspace}
\usepackage[nameinlink]{cleveref}
\usepackage[group-separator={,}, group-four-digits, per-mode=symbol, binary-units]{siunitx}

  \usepackage[caption=false,font=footnotesize]{subfig}
\ifCLASSINFOpdf
\else
\fi

\begin{document}
%TC:ignore
\title{Reconfigurable Intelligent Surface for Physical Layer Key Generation: Constructive or Destructive?}

\author{Guyue Li,~\IEEEmembership{Member,~IEEE}, Lei Hu,~\IEEEmembership{Student Member,~IEEE}, Paul Staat, Harald Elders-Boll, \\ Christian Zenger,~\IEEEmembership{Member,~IEEE}, Christof Paar,~\IEEEmembership{Fellow,~IEEE}, and Aiqun Hu,~\IEEEmembership{Senior Member,~IEEE}

	\thanks{Guyue Li and Lei Hu are with the School of Cyber Science and Engineering, Southeast
	University, Nanjing 210096, China (e-mail: guyuelee@seu.edu.cn;lei-hu@seu.edu.cn).}
	\thanks{Paul Staat is with Max Planck Institute for Security and Privacy, Bochum, Germany and 
		TH Köln – University of Applied Sciences, Cologne, Germany (e-mail: paul.staat@csp.mpg.de).}
	\thanks{Harald Elders-Boll is with TH Köln – University of Applied Sciences, Cologne, Germany  (e-mail: harald.elders-boll@th-koeln.de).}
	\thanks{Christian Zenger is with PHYSEC GmbH, Bochum, Germany and Ruhr University Bochum, Germany  (e-mail: christian.zenger@rub.de).}
	\thanks{Christof Paar is with Max Planck Institute for Security and Privacy, Bochum, Germany  (e-mail: christof.paar@csp.mpg.de).}
	\thanks{Aiqun Hu is with the National Mobile Communications Research Laboratory, Southeast University, Nanjing 210096, China (e-mail: aqhu@seu.edu.cn).}
	\thanks{Guyue Li and Aiqun Hu are also with the Purple Mountain Laboratories for Network and Communication Security, Nanjing 210096, China.
	 }
 }
\maketitle
%\IEEEtitleabstractindextext{%
\begin{abstract}
Physical layer key generation (PKG) is a promising means to provide on-the-fly shared secret keys by exploiting the intrinsic randomness of the radio channel. However, the performance of PKG is highly dependent on the propagation environments. Due to its feature of controlling the wireless environment, reconfigurable intelligent surface~(RIS) is appealing to be applied in PKG. In this paper, in contrast to the existing literature, we investigate both the constructive and destructive effects of RIS on the PKG scheme. For the constructive aspect, we have identified static and wave-blockage environments as two RIS-empowered-PKG applications in future wireless systems. In particular, our experimental results in a static environment showed that RIS can enhance the entropy of the secret key, achieving a key generation rate (KGR) of 97.39 bit/s with a bit disagreement rate (BDR) of 0.083. In multi-user systems where some remote users are in worse channel conditions, the proposed RIS-assisted PKG  algorithm improves the sum secret key rate by more than 2 dB, compared to the literature. Furthermore, we point out that RIS could be utilized by an attacker to perform new jamming and leakage attacks and give countermeasures, respectively. Finally, we outline future research directions for PKG systems in light of the RIS.
\end{abstract}

%TC:endignore
%}

%
%\IEEEdisplaynontitleabstractindextext
%
%\IEEEpeerreviewmaketitle

\section{Introduction} \label{sec:introduction}
Physical-layer key generation (PKG) is a promising approach to establish symmetric keys without complex ciphers in the ubiquitously connected wireless communication networks. By exploiting the intrinsic randomness of the radio channel, PKG allows two parties to generate shared secret keys from estimates of the wireless channels between them in a plug-and-play manner~\cite{2021Encrypting}.  
Nevertheless, despite its attractive advantages, PKG has an inherent deficiency that it needs a rich-scattering and dynamically changing channel to meet the requirements of consistency and security for the secret keys~\cite{AR}. Therefore, it is difficult for PKG to establish desired secret keys in some harsh propagation environments where the endpoints have low mobility or low signal-to-noise ratio (SNR).

Previous research works addressed this challenge by employing a cooperative relay node~\cite{relay}. However, the relay node, who could be untrusted, knows partial or even all information of the secret key. Although the information leakage could be alleviated by receiver-transmitted artificial noise~\cite{AR}, the legitimate receiver faces difficulties of self-interference elimination in practice. 
In addition, the secret key generation rate has limited growth unless the relay node keeps moving all the time. 
To further boost the rate, artificial randomness (AR) was introduced to form a fast-changing virtual channel that is a product of the physical channel and local AR at transceivers~\cite{AR}. However, the AR-aided PKG approach needs to modify the pilot signal, which is challenging to be implemented on commercial off-the-shelf (COTS) devices, e.g., ZigBee~\cite{2021Encrypting}, Wi-Fi~\cite{Low-entropy}. 
These vivid examples highlight that PKG still faces challenges and thus there is a strong need to facilitate PKG in the respective scenarios.

Very recently, an emerging technique to address this need is referred to as reconfigurable intelligent surface (RIS), which is capable of creating an intelligent reconfigurable propagation environment. In particular, a RIS is a planar array of a large number of reconfigurable passive elements (e.g., low-cost printed dipoles)~\cite{pan2021reconfigurable}. 
Each of the elements is able to induce a certain phase shift independently on the incident signal, thus collaboratively changing the reflected signal propagation. 
As of yet, only a few works have investigated the constructive effect of RIS to boost secret key generation rate.
In~\cite{Jirandom}, a RIS with a random phase shifting scheme was used to induce virtual fast fading channels for key generation and thus support one-time pad (OTP) encrypted data transmission with high rate. Further, \cite{SPletter} and \cite{Ji} optimized the used RIS units and the reflecting coefficient of RIS units to maximize the secret key generation rate, respectively.  However, in these works, the role of RIS on PKG is more like a ``nice to have" rather than a ``must-have" one.
The killer applications of using RIS for PKG are not clear and how to integrate it with various PKG systems is largely open. In addition, RIS can take destructive as well as constructive roles, decided by who owns the control right.  In \cite{hu2021ris}, it has been shown that when a RIS is controlled by a malicious party, it could break the key agreement of legitimate parties. The potential destructive aspects of RIS are largely overlooked in existing works. How to detect and defeat these attacks is still unknown. 
%Due to its feature of controlling the wireless environment, RIS may be a suitable solution for the synthesis of dynamic channels with high entropy, which is appealing to be applied in the difficult scenarios of PKG, as described in Fig.~\ref{Scenarios}.

Although a number of surveys and tutorials have recently appeared on the topic of integrating RIS with various communication technologies, there are no tutorials that overview and study RIS and PKG combined.
%It should be noted that existing survey and tutorial papers on PKG do not consider the RIS, which has attractive advantages to address the limitations of PKG. 
A comprehensive study of RIS-involved PKG, identifying challenges and opportunities, is still lacked.
In summary, this article includes the following key contributions:
\begin{itemize}
\item The role that RIS plays in PKG systems is studied from both constructive and destructive aspects. Specifically, static and wave-blockage environments are introduced as two RIS-empowered-PKG applications in future wireless systems, while RIS jamming (RISJ) and RIS leakage (RISL) attacks are identified as two categories of RIS-based attacks against PKG.
\item The entropy of the secret key is shown to be largely enhanced by the random surface configuration of RIS in real indoor environments and the sum secret key rate among multiple users is optimized through the flexible control on the phase shifts of RIS in wave-blockage environments.
%The feasibility of RIS-assisted multi-user PKG systems is evaluated under different configuration algorithms.
%by providing a key generation rate (KGR) of 97.39 bit/s with a bit disagreement ratio (BDR) of $0.083$. 
\item   An attacker-controlled RIS is shown to be able to disrupt the key establishment process or to help attackers obtaining the extracted secret keys between Alice and Bob. 
The proposed countermeasure based on channel separation is shown to be effective against the RISJ attack in wideband systems, while the leakage in RISL will be alleviated through using dynamic private pilots or adding more cooperative RISs.
\end{itemize}

%\begin{figure}[t]
%	\centering
%	\includegraphics[width=3.3in]{figures/magazine-1.pdf}
%	\caption{Application scenarios for RIS-assisted key generation systems.}
%	\label{Scenarios}
%\end{figure}

%TC:ignore
%\begin{figure*}[!t]
%\centering
%%\hspace*{\fill}%
%\subfloat[]{%
%	\includegraphics[width=6.3in]{figures/scenarios-a.pdf}
%	%\label{fig:scenarios-a}
%	}
%\\
%\centering
%\subfloat[]{%
%\includegraphics[width=6.3in]{figures/scenarios-b.pdf}
%}
%%\hspace*{\fill}%
%\centering
%\caption{Application scenarios for RIS-assisted key generation systems: (a) wave-blockage environments; (b) static environments.}
%\label{Scenarios}
%\end{figure*}
%%TC:endignore

\begin{figure}[t]
	\centering
	\includegraphics[width=0.9\linewidth]{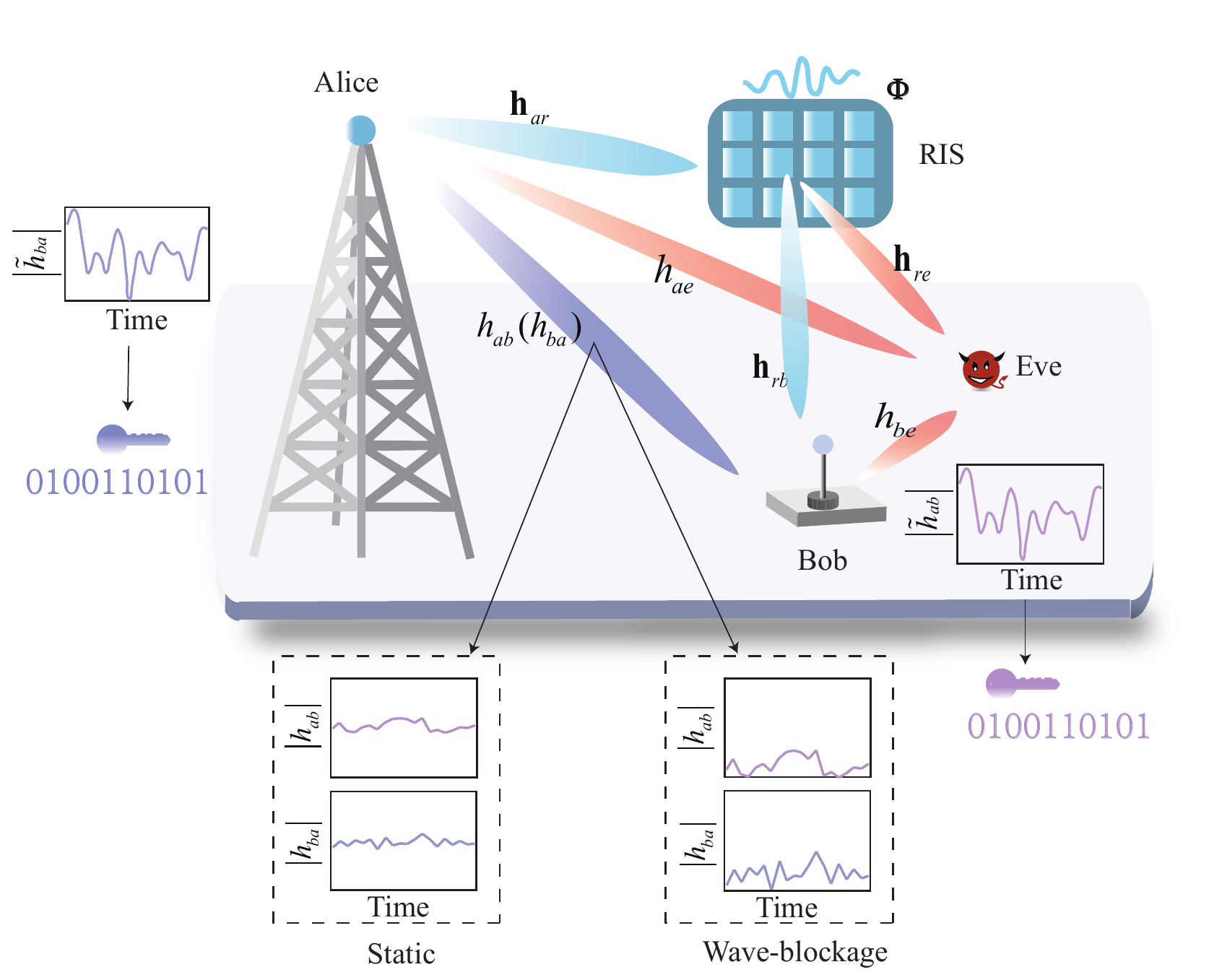}
	\caption{Application scenarios for RIS-assisted key generation systems.}
\label{Scenarios}
\end{figure}

\section{New Changes in Fundamentals When PKG Meets RIS}
\label{sec:pkg_meets_ris}
%In this section, we provide fundamentals on PKG, introduce the RIS-involved channel model, and discuss new changes when PKG meets RIS.
\subsection{Fundamentals of PKG}
%\subsubsection{The process of PKG}
In the model of a typical point-to-point PKG, there are generally three parties, where Alice and Bob are two legitimate parties who treat the wireless channel as a shared randomness source to extract a symmetric key, while Eve is a passive or active attacker who intends to eavesdrop the key or to prevent Alice and Bob from agreeing on the same key. 
%More legitimate and malicious parties are considered in the models of key generation among multiple users.
The general process of PKG comprises the following four phases.
\begin{itemize}
	\item \textit{Channel sounding:} In a time division duplex (TDD) system, Alice and Bob exchange pilots alternatingly to obtain bidirectional estimates of the channel between them and extract appropriate channel features, e.g., received signal strength~(RSS) and channel state information~(CSI). 
	\item \textit{Quantization:} Channel features are mitigated from disadvantageous stochastic properties (with respect to entropy as well as to reciprocity robustness) and converted into binary bit sequences, which are referred to as raw keys.
	\item \textit{Information reconciliation:} Possible disagreements between the two sequences are corrected via error-detection protocols or error-correcting codes.
	\item \textit{Privacy amplification:} Alice and Bob distill the key and wipe out possible information leakage from the previous phases.
\end{itemize}
As observed from the process, the feasibility of the PKG method is strongly associated with the wireless channel between Alice and Bob. %In particular, its reliability relies on the similarity of the bidirectional channels, while its security is ensured by the inherent randomness of the channel as well as the spatial diversity related information insufficiency of the attacker. 

%\subsubsection{The characteristics of RIS}
%Recently, RIS has received attention from the PKG community, as it has the ability to dynamically change the wireless channel between Alice and Bob. 
%We here review the basic architecture of RIS to see how it works. The structural design of RIS is not unified and a variety of design methods of RIS are given in the available literature. Despite that, t
%The architecture of an RIS usually consists of three layers, where the first layer is composed of numerous metallic elements to reflect incident signals, the second layer dynamically adjusts the reflection coefficient and the third layer is employed to receive control signals from other devices. 
%
%The implementation of RIS element is typically based on the PIN diodes. Specifically, the phase shift difference of $\pi$ in the incident signal can be realized by switching the "On" and "Off" states of the PIN diode~\cite{Tutorial}. 
%Ideally, when each element integrates a large number of PIN diodes, the amplitude and phase of each RIS are approximately continuous and independently adjustable, providing a theoretical performance limit. However, in practice due to the hardware cost and the limit of the number of control bits, the amplitude and phase shift of each element only have finite discrete values, e.g. power on/off control~\cite{SPletter}.

\subsection{RIS-involved Channel Model}\label{sec:model}
When a RIS participates in the process of PKG, either constructively or destructively, appropriate channel models must be found. %
%Moreover, the RIS-involved channel model changes fundamental principles of PKG, which should be taken into account in the system design.
%it will make difference to the channel model principles and hereby contributes to new theoretical limitations of the secret key rate.
%\subsubsection{Channel Model} 
On the basis of traditional channel models, the RIS introduces an additional channel between Alice and Bob. The RIS-induced channel conventionally includes the channel responses vector from single-antenna user Alice to the RIS, ${\bf h}_{ar}$, a diagonal matrix that models the RIS’ signal reflection, ${\bf \Phi_1}$, and the channel responses vector from the RIS to single-antenna user Bob, ${\bf h}_{rb}$. 
Specifically, the RIS receives the pilot signal from Alice and then reflects the signal while applying amplitude and phase changes adjusted by the RIS control layer. 
As a result, the Alice-RIS-Bob link, $\tilde h_{ab}$, is represented by a multiplicative channel model, which is added coherently with the direct link, $h_{ab}$, to form the new channel model as
\begin{align} 
\tilde h_{ab}= {\bf h}_{ar}^T{\bf \Phi_1}{\bf h}_{rb}+h_{ab}.
\end{align} %For a practical channel model, propagation phenomena including, i.e., far-field effect, mutual coupling, path loss, and multi-path, should be all incorporated in these channel matrices.
Similarly, the reverse channel from Bob to Alice, $\tilde h_{ba}$, satisfies that 
\begin{align}
\tilde h_{ba}= {\bf h}_{br}^T{\bf \Phi_2}{\bf h}_{ra}+h_{ba},
\end{align}
where ${\bf h}_{br}$, ${\bf h}_{ra}$, and $h_{ba}$ respectively denote the channels from Bob to the RIS, from the RIS to Alice, and from Bob to Alice, and ${\bf \Phi_2}$ denotes the reflection matrix of the RIS in the reverse channel. The channels observed by Eve also follow the above model by substituting the direct channel with the channel from Alice/Bob to Eve and substituting the channel responses from the RIS to Bob/Alice in the RIS-induced channel with those from the RIS to Eve.

%as $\tilde h_{ae}={\bf h}_{ar}^T{\bf \Phi_1}{\bf h}_{re}+h_{ae}$ and $\tilde h_{be}={\bf h}_{br}^T{\bf \Phi_2}{\bf h}_{re}+h_{be}$, where ${\bf h}_{re}$, $h_{ae}$ and $h_{be}$ denote the channels from the RIS, Alice and Bob to Eve, respectively. 
%\subsubsection{Principles} 

The reflection coefficients of RIS elements in ${\bf \Phi_1}$ and ${\bf \Phi_2}$ are often programmed to achieve purposeful manipulation of wireless channel. Ideally, these reflection coefficients, including phase shifts and amplitudes, are assumed to be continuously tunable to achieve the optimal performance, which provides a theoretical bound for RIS-assisted systems. However, considering the hardware cost and implementation complexity, some works assume the set of reflection coefficients to be discrete with finite amplitude or phase shift levels. 
For example, in a two-level phase shift control, the reflection coefficient of an element may be controlled by a PIN diode~\cite{pan2021reconfigurable}, selecting from $\{1,-1\}$, i.e., by switching biasing ``On" and ``Off".

\subsection{New Changes} 
Since the fundamental principles of PKG, i.e., channel reciprocity, spatial diversity, and temporal fluctuation, depend on the channel model, an interesting question is whether these principles will still be met under the new RIS-involved channel model. 
\begin{itemize}
	\item \textit{Channel Reciprocity:} The reciprocity between $\tilde h_{ab}$ and $\tilde h_{ba}$ is not merely dependent on the reciprocity of the physical channels. It also depends on the agreement of the diagonal matrices ${\bf \Phi_1}$ and ${\bf \Phi_2}$ in the bidirectional channel probing phases. Besides, for the channel features, the non-reciprocity effect of noise is changed as the received signal strength can be either boosted or attenuated through the RIS.
	\item \textit{Spatial Diversity:} The correlation between channel features of Bob and Eve is also changed by RIS. According to the central limit theorem, when the number of reflecting elements is large, the RIS-induced channels between Bob and Eve asymptotically satisfy the independence as long as the RIS reflection channels of Eve and legitimate parties are independent~\cite{sum}.
	\item \textit{Temporal Fluctuation:} RIS introduces an additional degree of freedom, which is capable to boost the randomness. Through a dynamical configuration of RIS, the channels of $\tilde h_{ab}$ and  $\tilde h_{ba}$ vary along time, even for environments where the physical channel would not fluctuate strongly.
\end{itemize}
Accordingly, when the RIS is controlled by legitimate users, it is possible for them to meet these principles, even in some harsh scenarios.
Conversely, RIS can also be exploited by Eve to disrupt the secret key generation between Alice and Bob by breaking these principles. %The constructive and destructive effects of RIS on PKG systems will be elaborated in Sec.~\ref{Sec:III} and Sec.~\ref{Sec:IV}, respectively.

%\subsubsection{Secret Key Rate}
%Traditionally, the secret key rate is defined as the conditional mutual information of \blue{CZ: raw?} channel observations of Alice and Bob, given the observation of Eve \cite{zeng}. %, i.e. $R = I(h_A;h_B|h_E)$. It represents the theoretical limitation of a PKG method and provides an objective function in the optimal design of RIS phase shifts. Since RIS makes the channel model more complex, the distributions of channel features are generally difficult to calculate and  thus it is challenging to derive a closed-form expression for the secret key rate. Fortunately, according to the central limit theorem, the distribution of the RIS-induced channel approximates a normal distribution when the RIS element number becomes larger\footnote{Theoretically, when the number of RIS elements is larger than $16$, the probability distribution function (PDF) is well approximated by a complex Gaussian distribution.}.  Based on this assumption, the close-form of secret key rate is derived in \cite{chen2021intelligent} and \cite{Ji} for the cases of single antenna and multiple antennas, respectively.
\section{RIS-assisted PKG: Potential Improvements}
\label{Sec:III}
%Due to its benefits, the RIS is ideal as a helper, assisting Alice and Bob to form an RIS-induced fluctuating channel, which serves as the common randomness for key generation. 
%In this section, we analyze typical scenarios in which RIS assists PKG and summarize the existing RIS reflecting coefficient design methods in the channel sounding stage. We conclude the section with initial experimental results, demonstrating RIS-assisted PKG in a practical setting. 
%In this section, we  analyze the constructive effects of RIS by considering two case studies on RIS-assisted PKG to tackle static and wave-blockage environments, respectively.
In this section, we analyze the constructive effects of RIS by considering two case studies with ``must-have" RIS environments, as shown in Fig.~\ref{Scenarios}. %\blue{Apart from the given two cases, RIS plays a ``nice to have'' role to further improve the secret key rate, which have been studied in existing works \cite{Ji,SPletter}.

\begin{figure}[t]
	\centering
	\includegraphics[width=0.99\linewidth]{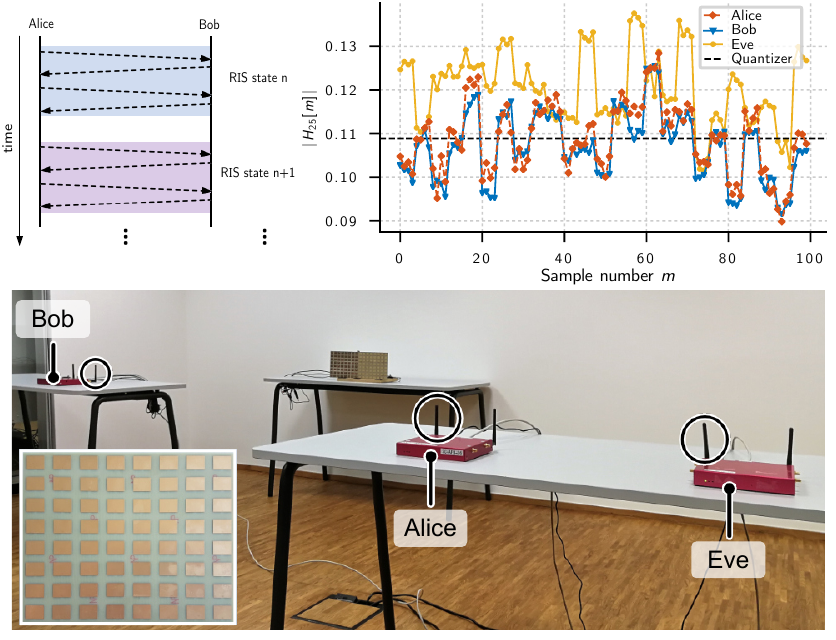}
	\caption{Experimental setup, RIS-assisted PKG protocol and time-domain signals resulting from channel probing by Alice, Bob, and Eve on a single OFDM subcarrier. The magnified view to the lower left shows the front view of a RIS module with reflector patch elements (\SI{20}{cm} x \SI{16}{cm}).}
    \label{fig:irs_photo_setup}
\end{figure}

\subsection{Case I: Static environments}
%The straightforward and opportunistic nature of PKG promotes investigations in field studies and considerable previous research activities in this area resulted in today's standard PKG architectures~\cite{var}.%~\cite{aono_wireless_2005} 
%innovative solutions have been proposed for introducing artificial randomness or using reconfigurable antennas.
In static environments, the limited randomness found in the direct channel renders key generation challenging, while RIS-induced channel variations can now provide the basis for PKG.
%Since the security of PKG depends on the randomness of the wireless channel, in the quasi-static environment where the channel remains unchanged for a long time, the generated key has a long string of zeros or ones. 
%At this point, the generated key has low entropy and an attacker can crack the key through a dictionary attack. 
%\blue{Since RIS consists of reflective elements that can be independently adjusted, it can be used as a randomness source to increase the changes of the wireless environment~\cite{Low-entropy,Jirandom}}. 
%Random phase-based IRS has been proposed to increase channel time-variability and achieve one-time-pad encryption \cite{Jin,Jirandom}. 
%More channel randomness will be obtained when the number of RIS elements is large. 
Thus, with the help of RIS in PKG, the entropy of the secret key can be enhanced~\cite{Low-entropy,Jirandom}. 
% \blue{Compared with existing AR and relay assisted PKG Methods, RIS can match existing communication protocols and has lower hardware cost and energy consumption.}

Here, we report promising experimental results of a RIS-assisted PKG system in static environments.
\subsubsection{Setup}
\label{sec:experimental_setup}

%\begin{figure*}
%\hspace*{\fill}%
%\subfloat[]{%
%	\includegraphics[width=0.3\linewidth]{figures/fig_irs_sampling.pdf}}
%\hfill
%\subfloat[]{%
%\includegraphics[width=0.8\columnwidth]{figures/td_journal_64.pdf}
%}
%\hspace*{\fill}%
%\caption{(a) RIS-assisted PKG protocol, two packet exchanges per RIS state. (b) Time-domain signals resulting from channel probing by Alice and Bob on a single OFDM subcarrier.}
%\label{fig:td_channel_probing}
%\end{figure*}
%% Alternative presentation using subfigure %%%

% \begin{figure}
% \hspace*{\fill}%
% \subfloat[]{%
%        \includegraphics[width=0.4\columnwidth]{figures/irs_front_cut_perspective.jpg}}
% \hfill
% \subfloat[]{%
% \includegraphics[width=0.56\columnwidth]{figures/IMG_20210630_214019_edit-scaled.jpg}
% }
% \hspace*{\fill}%
% \caption{Intelligent reflecting surface prototype module. (a) Front view with patch elements (\SI{20}{cm} x \SI{16}{cm}). (b) Experimental setup.}
% \label{fig:irs_photo_phase}
% \end{figure}
For our field trials, we place the parties Alice and Bob as well as a RIS in a long-term static basement environment.
%, which until now would have been inadequate for PKG due to the lack of channel variation. However, RIS-induced channel variations can now provide the basis for PKG.
We use commodity \mbox{Wi-Fi} devices, implementing orthogonal frequeny division multiplexing~(OFDM) communication at Alice and Bob, in conjunction with a RIS prototype to realize a PKG system as outlined in Section~\ref{sec:pkg_meets_ris}. For the channel probing, we employ a C application to control ath9k-based PCIe network interface cards~(NICs) implementing IEEE 802.11n Wi-Fi in a 2x2 MIMO configuration. The RIS consists of two modules (see Fig.~\ref{fig:irs_photo_setup}) with a total of $128$ elements, each having $1$-bit phase control, i.\@\,e.\@\xspace, an impinging wave is either reflected with phase shift $0$ or $\pi$. The elements are switched via a micro-controller using serial communication. %$0^\circ$ or $180^\circ$.
To generate the desired channel variation, the RIS applies random surface configurations at a fixed update rate. In parallel, Alice and Bob exchange \mbox{Wi-Fi} packets to obtain CSI in the channel probing phase. We use \mbox{Wi-Fi} channel 60 at 5.3~GHz which is close to the RIS' optimum operation frequency at 5.37~GHz. Alice and Bob are \SI{3}{\m} and \SI{1.5}{\m} apart from the RIS, with line-of-sight for the direct and the RIS channels.

%The effective channel between Alice and Bob, $H_k[n]$ with the $k^{th}$~OFDM subcarrier for $n^{th}$~packet, is the sum of the $128$ cascaded sub-channels via the IRS and the direct (non-IRS) channel. Thus, we may denote $H_{k}[n] = \sum_{i=0}^{128} h_{i,k}\, r_i[n]\, g_{i,k} + d_k$, where $h_{i,k}, g_{i,k}, d_k \in \mathbb{C}, r_i \in \{-1,1\}$, respectively, are the complex channel gains of the link between Alice and the $i^{th}$ IRS element, Bob and the $i^{th}$ IRS element, the direct link, and the reflection coefficient $r_i$. With random IRS configurations $r_i$, according to the central limit theorem, we expect $H_k[n]$ to converge to a complex normal distribution, well-suited for PKG.

\subsubsection{Results}
%TC:ignore

%\begin{figure}[t]
%	\centering
%	\includegraphics[width=0.8\linewidth]{figures/td_journal_64.pdf}
%	\caption{Time-domain signals resulting from channel probing by Alice and Bob on a single OFDM subcarrier.}
%    \label{fig:td_channel_probing}
%\end{figure}
%TC:endignore
The channel probing phase and the respective RIS states are illustrated at the top left of Fig.~\ref{fig:irs_photo_setup}. We consider uncoordinated RIS operation where Alice and Bob adapt their channel probing to the speed at which the channel varies, as is the case with conventional PKG implementations. Due to the switched RIS behavior, the corresponding channel component changes instantaneously and thus Alice and Bob may \textit{oversample} the channel response in between, i.\@\,e.\@\xspace, spend $L$~channel measurements per RIS configuration. Exemplary for the instantaneous RIS-induced channel variation and oversampling ($L=4$), we plot the magnitude CSI of Alice and Bob for a single subcarrier~$k=25$ at the top right of Fig.~\ref{fig:irs_photo_setup}. After block averaging over $L$ samples to improve SNR, both parties translate their respective channel profiles to bit strings using single bit CDF quantization~\cite{quantizer}. In our particular implementation, one bidirectional packet exchange for channel probing takes $T_p \approx \SI{2}{\ms}$ and we consider a RIS configuration duration of $T_{RIS} = L\, T_p$. Accounting for an additional RIS update time $T_{u} \approx \SI{2}{\ms}$, we can estimate the effective key generation rate~(KGR) after quantization %for single bit quantization
%\blue{CZ: How do we calculate the KGR (definition is missing)? Is it done also after quantization? Do we use the NIST results (or entropy estimation) to calculate the KGR? What is the differece to the secret key rate? It is usually $R_k = I(X_A;Y_B)-max{I(X_A;Z_E);I(Y_B;Z_E)}$. Do we calculate the mutual information $I(X;Y)$? Ho are the results of teh simulation (Secret key rate) comparable with results of the experiments (KGR)?}
with $(L\, T_p + T_{u})^{-1}$. This is consistent with the observation that the RIS here dictates the channel coherence time, which in turn allows to formulate an upper bound on the KGR. Table~\ref{tab:rates} shows the expected and experimentally achieved KGR and bit disagreement rate (BDR) results for an exemplary OFDM subcarrier $k=25$ with varying values of $L$ with and without the RIS. For the latter, the BDR of approximately \SI{50}{\percent} highlights the infeasibility of PKG in static environments.

%TC:ignore
\begin{table}[ht]
\centering
\caption{Single subcarrier KGR and BDR}
\label{tab:rates}
\begin{tabular}{@{}lrccccc@{}}
\toprule
%& \multicolumn{4}{c}{Country List} \\
&                    & \textbf{$L=1$}    & \textbf{$L=2$} & \textbf{$L=3$} & \textbf{$L=4$}\\
\toprule
%\textbf{KGR [bit/s]}                    & \\
\multirow{2}{*}{\textbf{KGR [bit/s]}}& Expected & 250.00   & 166.67   & 125.00  &  100.00 \\
& Measured & 237.45   & 160.58   & 121.30  &  97.39 \\
\midrule 
\multirow{2}{*}{\textbf{BDR, $k=25$}} & With RIS                   & 0.260   & 0.157   & 0.108  & 0.083  \\
%BDR, $k=94$                   & 0.218   & 0.120  & 0.082  &  0.065 %\\
& Without RIS            & 0.481   & 0.492  & 0.491  & 0.488  \\
%BDR, $k=94$                   & 0.499   & 0.489  & 0.486  & 0.495\\
\bottomrule 
\end{tabular}
\end{table}
%TC:endignore
%\blue{CZ: with respect to the table: Could we calculate the secret key rate instead (or at least the BDR between Bob and Eve)? }
We assess the randomness of the key material generated by the RIS-assisted PKG procedure. Therefore, we apply the statistical test suite for random number generators provided by the American NIST~\cite{rukhin_statistical_2010}. We examined a bit sequence of length \SI{600000}{bits} that was obtained at the output of the quantization stage without further processing. All applicable tests were passed, indicating that the randomized RIS-induced channel variations are suitable for cryptographic applications. %We stress that the RIS provides the entropy for the key material and therefore requires a high-quality randomness source, e.\@\,g.\@\xspace, a true random number generator~(TRNG).

In another experiment, we placed an eavesdropper Eve at \SI{1}{\m} distance to Alice. Here, the security of the key material is linked to the correlation of the channels from the RIS to Alice and Eve, respectively. We estimate the mutual information in the raw channel sequences of \SI{100000}~random RIS configurations observed by Alice, Bob, and Eve for a single OFDM subcarrier. To estimate the empirical mutual information, we utilize a $k$-nearest neighbor estimation~\cite{kraskovEstimatingMutualInformation2004}. We found the average mutual information across OFDM subcarriers of Alice' and Bob's and Alice' and Eve's observations to be $0.69$ and $0.04$ bit/observation, respectively. This shows that Eve is at a significant disadvantage compared to the legitimate parties.

%In the experiment, we found average Pearson correlation coefficients across OFDM subcarriers of $0.74$ for Alice and Bob and $0.18$ for Eve. 
%\blue{CZ: could we calculate the mutual information (and/or BDRs) instead? It would be more comparable with the simulation results.} 

Our results highlight the RIS as a convenient extension to PKG, enabling key generation in static environments. As the RIS is a part of the environment, operating on directly on the physical layer and external to user terminals, all users within reach -- not just Alice and Bob -- can potentially benefit from the provided channel randomization effect. As demonstrated, existing devices, wireless standards, and PKG procedures~(cf.~Section~\ref{sec:pkg_meets_ris}) can be reused with little overhead. %Beyond the demonstrated use-case, RIS-assisted PKG offers much potential for future work, e.\@\,g.\@\xspace, on optimized surface configurations with SNR enhancement and reduction of information leakage  to eavesdroppers. 
%\begin{figure*}
%	\hspace*{\fill}%
%	\subfloat[]{%
%		\includegraphics[width=0.4\linewidth]{figures/M1_1.eps}}
%	\hfill
%	\subfloat[]{%
%		\includegraphics[width=0.4\linewidth]{figures/M2.eps}
%	}
%	\hspace*{\fill}%
%	\caption{The sum secret key rates of RIS-assisted multi-user PKG systems. (a) Comparison for different RIS configurations versus the number of RIS elements under the assumption of independent channels among UTs. The number of users is four. (b) Comparison for different RIS configurations under assumptions of independent and correlated
%channels among UTs. The number of RIS elements is $16$ and the correlation coefficient is set to be $0.5$ for the simulation of correlated channels.}
%	\label{Fig:sum}
%\end{figure*}

\subsection{Case II: Wave-blockage environments}
In wave-blockage environments, the direct link is blocked, leading to low SNR at the receiver. Consequently, CSI is submerged in noise and thus results in a low BDR, which is the upper bound on the number of bits per channel observation that Alice and Bob can generate, about which Eve cannot obtain any useful information based on her own observation~\cite{var}. 
%\blue{CZ: I guess the BDR is calculated after quantization? What kind of quantization scheme is used? One-threshold/Single bit as shown in Figure 4b? Coule we cite the used quatizer? [https://www.emsec.ruhr-uni-bochum.de/research/publications/security-analysis-quantization-schemes-journal/] This should be clarified. --- As well as, why are the simulation results and the experimental data comparable?}
To facilitate PKG in wave-blockage environments, RIS shapes a RIS-induced fluctuating channel, which serves as the common randomness for generating secret keys. It has been shown in the literature that the user's SNR can be significantly increased by designing the configuration of a RIS appropriately~\cite{pan2021reconfigurable}. As an extension, in multi-user systems where some remote users are in worse channel conditions, the channel reciprocity of these users can be improved by the deployment of RISs. 
%Therefore, with a sophisticated design of the configuration, RIS is hopeful to help Alice and Bob to achieve a good channel reciprocity and a high secret key rate. 
Since the key generation performance, in this context, depends on the RIS configuration, some works~\cite{Ji,SPletter,sum} have explored the optimization of RIS configurations to achieve the maximum secret key rate.

In the former case, we have demonstrated the constructive effect of a RIS with random phase shifts on a single user PKG system in a static environment.
On this basis, we further report promising theoretical results of a RIS-assisted multi-user PKG system with different RIS configuration algorithms in wave-blockage environments.
\subsubsection{Setup}
We consider a RIS-assisted multiuser key generation system, which comprises a wireless AP (Alice), a RIS, an eavesdropper (Eve), and multiple legitimate user terminals (UTs). All parties, including Alice, Eve, and the UTs, are assumed to be equipped with a single antenna. Alice intends to generate individual secret keys with the UTs. The direct wireless channels between Alice and the UTs are blocked, therefore a RIS is deployed to enable the key generation. The phase shifts of the RIS are programmed and reconfigured via a controller. Eve is assumed to be a passive eavesdropper, who is located at least several wavelengths away from the legitimate parties. In the work of~\cite{sum}, we have proved that when the number of antenna elements of the RIS is large, the RIS-induced channels of Eve are independent of those of Alice and the UTs, therefore they do not affect the secret key rate. However, when Alice communicates with a UT, the other UTs are regarded as potential non-colluding curious users. The channel from Alice to the RIS and the channel from the RIS to each UT are zero-mean
complex Gaussian random variables. %Unless otherwise stated, the number of RIS elements is $16$, the number of UTs is four, and the SNR is $10$ dB, where the SNR is defined as the ratio of the transmitted power and the noise variance.
\subsubsection{Results}
In the simulation results, we evaluate the feasibility of RIS-assisted multi-user PKG systems through the metrics of sum secret key rate, which is defined as the sum of  secret key rates between Alice and all UTs. The secret key rate between Alice and one UT is expressed as the minimum mutual information given the observation of the other UTs. 
For traditional PKG systems without RIS, the sum secret key rate is zero, as the direct wireless channels between Alice and the UTs are blocked. For RIS-assisted multi-user PKG systems, we compare the sum secret key rates of them under three different configuration algorithms of RIS: (1) random phase shifts, (2) on-off switching states, and (3) optimized phase shifts. 
In the second algorithm~\cite{SPletter}, the reflection coefficients are in a two-level amplitude control and the secret key rate is maximized by turning on limited number of RIS units, which correspond to the largest variances of the RIS channels.  In last algorithm, we solve the optimal phase shift ranging from $-\pi$ to $\pi$ to maximize the sum secret key rate. 

\begin{figure}[t]
	\centering
	\includegraphics[width=0.8\linewidth]{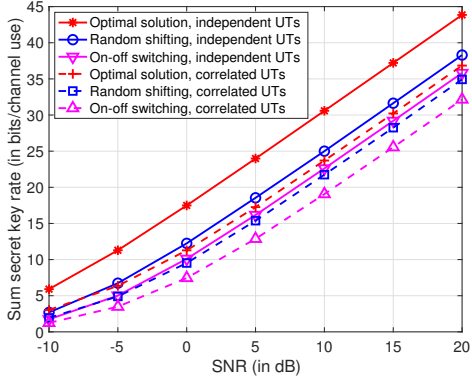}
	\caption{The comparison of sum secret key rates for different RIS configurations under assumptions of independent and correlated channels among UTs. The number of RIS elements is $16$ and the correlation coefficient is set to be $0.5$ for the simulation of correlated channels.}
    \label{Fig:sum}
\end{figure}
Fig.~\ref{Fig:sum} reports the sum secret key rates obtained by RIS-assisted multi-user PKG systems under these RIS configuration algorithms over independent and correlated channels\footnote{The correlated channels are realized by putting two independent channels through a correlation matrix.}. %Fig.~\ref{Fig:sum} (a) shows the sum secret key rate as a function of the number of RIS elements  under the assumption of independent channels among the UTs. 
%When reflection channels among the UTs are independent, the sum secrecy rates of the three considered configuration algorithms increase with the number of RIS elements and 
When the reflection channels among the different UTs are correlated, the RIS-induced channels are correlated as well, which causes information leakage among the UTs. It is observed that the spatial correlation among the channels of the UTs reduce the sum secret key rate for all of the considered RIS configuration algorithms. The performance loss is roughly $5$ dB for the optimal solution, $2$ dB and $4$ dB for the random configuration algorithm and the on-off switching algorithm, respectively. %The rate gap between independent and correlated channels grows slightly with SNR. 
For both independent and correlated channels, the optimal solution provides the best sum key rate, while the on-off switching algorithm has lower sum secret key rate than others due to its limited controllable states of RIS units. 
The performance gap between the optimal solution and the on-off switching algorithm is $7$ dB and $4$ dB for independent and correlated channels, respectively.
It should also be noted that although the optimal solution provides the best sum secret key rate, it has higher computation complexity than other two configuration algorithms. The other two configurations are also appealing for practical usage in systems with limited computing capacity.
%Fig.~\ref{Fig:sum} compares the sum secret key rate over independent and correlated channels as a function of the SNR. 

%\begin{figure*}
%    \hspace*{\fill}%
%    \subfloat[]{%
%        \includegraphics[width=0.25\linewidth]{figures/attack_model.pdf}}
%    \hfill
%    \subfloat[]{%
%        \includegraphics[width=0.7\columnwidth]{figures/BDR.pdf}
%    }
%    \hfill
%    \subfloat[]{%
%    \includegraphics[width=0.68\columnwidth]{figures/KGR.pdf}
%    }
%    \hspace*{\fill}%
%     \caption{The RISJ attack. (a) Attack model. (b) The BDR versus SNR under the RISJ attack in an OFDM system. (c) The KGR of the CCS-protected PKG method under different bandwidths.}
%    \label{fig:attack}
%\end{figure*}

\begin{figure}[t]
	\centering
	\includegraphics[width=0.8\linewidth]{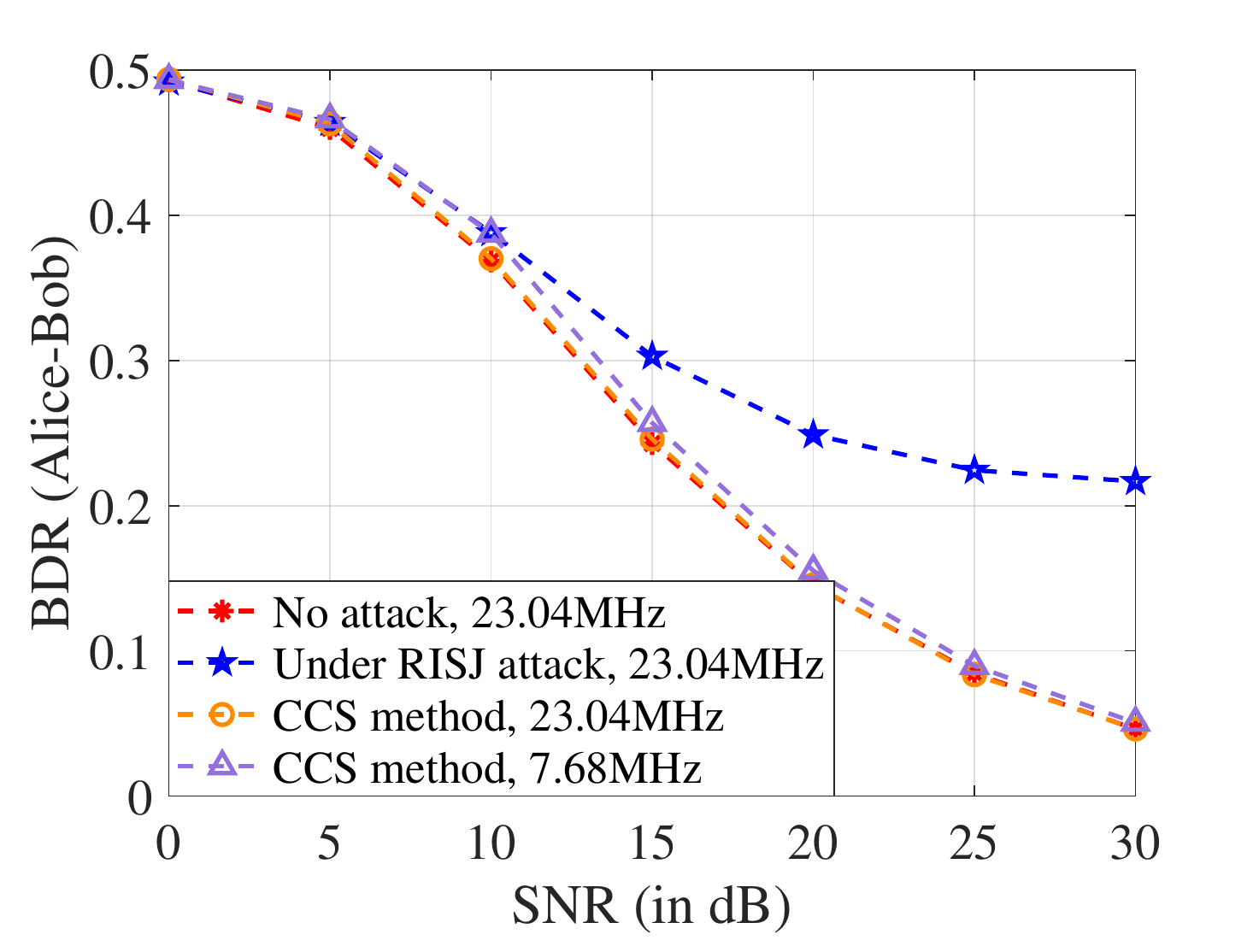}
	\caption{The BDR of Alice and Bob versus SNR under the RISJ attack and CCS-protected PKG method in an OFDM system.}
   \label{fig:attack1}
\end{figure}

% \begin{figure}[t]
% 	\centering
% 	\includegraphics[width=0.8\linewidth]{figures/KGR.pdf}
% 	\caption{The KGR of the CCS-protected PKG method in cases of no attack and under the RISJ attack in an OFDM system.}
%   \label{fig:attack}
% \end{figure}
%In more general scenarios, multiple RISs can be employed in conjunction to enhance SNR and increase channel variation to facilitate PKG. Moreover, in the presence of multiple eavesdroppers, the dense deployment of RIS will effectively inhibit eavesdropping and reduce information leakage. 

\section{RIS-BASED ATTACKS AND COUNTERMEASURES}
\label{Sec:IV}
Despite constructive effects as shown in Sec.~III, the RIS may also bring in destructive effects on PKG: Assuming an attacker-controlled RIS, a number of novel attacks against PKG become possible.
Based on the attacker's goal, we divide RIS-based attacks into two categories and discuss along countermeasures in this section.
\subsection{RIS Jamming (RISJ) attack}
\subsubsection{Attack Description}
RISJ attack is an active attack which aims at disrupting the key establishment process between Alice and Bob by adjusting the RIS reflection matrices. 
According to the channel model in Sec.~\ref{sec:model}, the RIS-involved channel is the superposition of the direct link and the RIS-induced link, therefore the channel reciprocity will be affected by either of them. This can be taken advantage of by Eve to launch a jamming-like attack through RIS in two ways.

First, Eve can change the RIS reflection matrices in the Alice-RIS-Bob link and the Bob-RIS-Alice link, i.e., ${\bf \Phi_1} \ne {\bf \Phi_2}$, in the process of a bidirectional channel probing. 
For example, when the RIS applies random surface configurations at an update rate higher than the channel sampling rate\footnote{For PIN diode-based RIS implementations, the switching frequency can be up to $5$~MHz, which corresponds to the switching time of $0.2$ $\mu$s~\cite{pan2021reconfigurable}. This is much smaller than the typical channel coherence times in the order milliseconds.}, the observed CSI at Alice and Bob will most likely to be different. 
% \blue{The severity of this attack increases with the energy proportion of the RIS-induced channel in the superposition, referred to as $\gamma$. Fig.~\ref{fig:attack1} reports  the BDR of raw keys generated from the CSI in an OFDM system under the RISJ attack with different $\gamma$. From the results, the BDR rises with the increase with $\gamma$. In particular, when $\gamma$ is $0.1$, the BDR reaches $0.25$ at the SNR of $20$ dB, which would impose a great burden to correct these errors. }
%\blue{Fig.~\ref{fig:attack1} reports the BDR of raw keys generated from the CSI in an OFDM system under the RISJ attack when the energy proportion of the RIS-induced channel $\gamma = 0.1$. }
Second, Eve can reduce the channel reciprocity by attenuating the received signal strength through the RIS, which is similar to traditional jamming attacks. This attack requires the RIS to have the ability of channel estimation, which might be endowed by the technology of Semi-Passive RIS~\cite{pan2021reconfigurable}.

Different from traditional jamming attacks, the RISJ attack reflects a jamming signal instead of transmitting it. 
Due to this fact, it is difficult to resist this attack by countermeasures based on location detection, since the location of Eve will be hardly exposed. Therefore, new ideas are desired by Alice and Bob to establish secret keys against the RISJ attack.
\subsubsection{Countermeasures}
One possibility is to exploit the fact that Eve can only change the RIS-induced channel, while the direct link, if exists, will not be affected by the RIS. Thus, Alice and Bob could separate them to distinguish the contaminated channel and still generate secret keys from the remaining uncontaminated channel. This idea, referred to as countermeasure based on channel separation (CCS), can be implemented in PKG systems with high multipath resolution.
For example, in a wideband OFDM system, these channels are separated through an inverse discrete Fourier transform (IDFT) and the RIS-induced channel can be detected from its high variance~\cite{hu2021ris}. 
Fig.~\ref{fig:attack1} reports the BDR of raw keys generated from the CSI in an OFDM system under the RISJ attack and CCS-protected PKG method when the energy proportion of the RIS-induced channel $\gamma = 0.1$. From the results, the BDR after RISJ attack is close to $0.2$ with the SNR increases, which would impose a great burden to correct these errors. However, the BDR of CCS-based method is almost the same with that of no attack at bandwidth of $23.04$ MHz, which verifies that CCS can resist the RISJ attack in a wideband system. When the bandwidth is $7.68$ MHz, there is a slight performance loss under the RISJ attack, which is caused by possible false alarm of the contaminated channel due to the limited multipath resolution. Moreover, the KGR of the CCS method in cases of no attack and under the RISJ attack are both 35 bits/channel use at bandwidth of $23.04$ MHz, and the performance loss is less than $20\%$ under the RISJ attack at bandwidth of $7.68$ MHz.
%Fig.~\ref{fig:attack} shows the KGR of the CCS-protected PKG method in cases of no attack and under the RISJ attack in an OFDM system.

%The KGRs are almost the same for both cases at bandwidth of $23.04$ MHz, which verifies that CCS can resist the RISJ attack in a wideband system. When the bandwidth is $7.68$ MHz, the performance loss is less than $20\%$ under the RISJ attack, which is caused by possible false alarm of the contaminated channel due to the limited multipath resolution.

However, it is still challenging to resist the RISJ attack for systems with low multipath resolution, e.g., narrow-band systems. One solution is to reduce the energy proportion of the contaminated RIS channel, e.g., by introducing more RIS devices as helpers.

\subsection{RIS Leakage (RISL) attack}
\subsubsection{Attack description}
In a RISL attack, Eve is more interested in obtaining the extracted secret keys between Alice and Bob, but not preventing them from agreeing on the same key. Given the fact that the RIS-induced channel, as a component of the channel for key generation, is under control by Eve, he/she is likely to obtain partial or complete information of the secret key. Here, we introduce two kinds of RISL attacks. 

The first RISL type intends to cause desired or predicable changes in the channel measurements by changing the RIS reflection matrices in a planned way.
This RISL attack is similar to the well-known predictable channel attack, but one clear difference between them lies in the manipulation object.
Compared with controlling the movements of some intermediate object, manipulating RIS reflection matrices is more undetectable and easier to implement.
For example, when Eve switches the ``On” and ``Off” states of all elements on the RIS alternatingly, the channel gain observed at Alice and Bob is likely to be boosted and attenuated regularly with time. These channel values would lead to an alternating sequence of $0$ and $1$ in raw keys. Such predictable keys are easy to break. We have performed this attack using the experimental setup from Section~\ref{sec:experimental_setup}. While the BDR of Alice and Bob is at \SI{6.5}{\percent}, the BDR of Eve and Alice is only \SI{5.8}{\percent}.

In the second RISL attack, Eve speculates on the legitimate channel measurement by calculating the gain of the RIS-induced channel from the multiplicative channel model.
In this attack, Eve needs to accurately estimate the channel gains from Alice and Bob to RIS, which might impose a high requirement for Eve.
Since Eve can hardly obtain the channel information of the direct link, these RISL attacks would be more effective in some harsh propagation environments, e.g., static environments and wave-blockage environments. 
\subsubsection{Countermeasures}
\begin{figure}[t]
	\centering
	\includegraphics[width=0.8\linewidth]{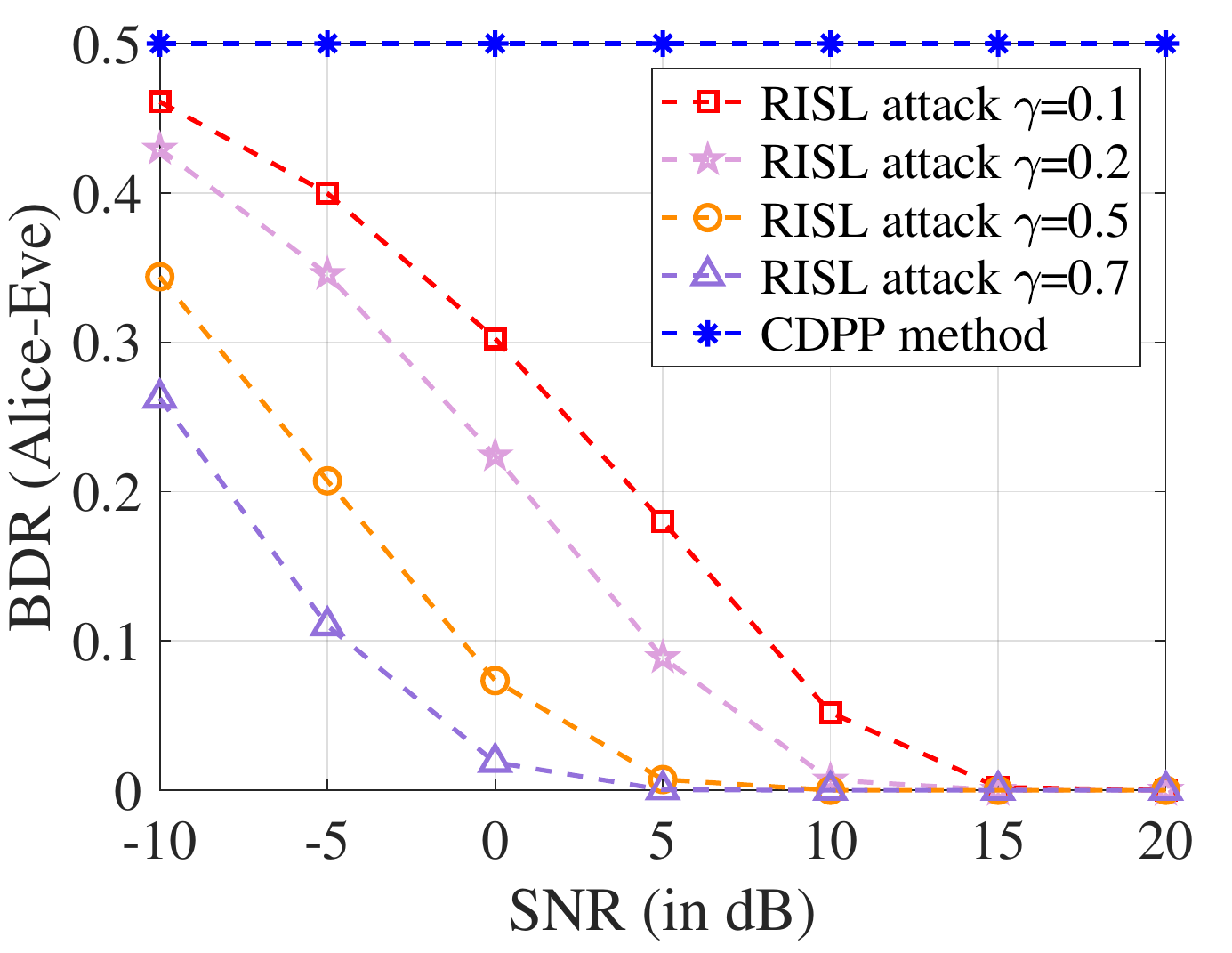}
	\caption{The BDR of Alice and Eve under RISL attack and CDPP-protected PKG method in static environments.}
   \label{fig:RISL_BDR}
\end{figure}

The effect of RISL is limited in fast-varying and rich multipath scattering environments, therefore we focus on countermeasures in static and  wave-blockage environments, respectively. In a static environment, RISL attacks can be prevented by increasing the fluctuation of the direct link.
With the help of dynamic private pilots or artificial noise, Alice and Bob could construct an artificial fast-fading channel.
Since the channel fluctuation is not completely controlled by the RIS, it is difficult for Eve to obtain the secret keys, either by predicable changes or by channel speculation. 
Fig.~\ref{fig:RISL_BDR} illustrates the simulation results of RISL attack and the countermeasure based on dynamic private pilots (CDPP), where single threshold quantization method based on RSS is employed. From the results, the BDR of Alice and Eve is observed to reduce with the increase of $\gamma$ and SNR. Specifically, in the case of $\gamma = 0.2$ and SNR = 10 dB, the bits of Alice are completely obtained by Eve.
%\begin{figure}[t]
%		\centering
%		{\includegraphics[width=0.8\textwidth]{fig/RISL_BDR.pdf}}
%		\caption{The BDR of Alice and Eve under RISL attack and corresponding countermeasure in static environments.}
%		\label{fig:RISL_BDR}
%\end{figure}
In our proposed CDPP method, under different $\gamma$ and SNR, the BDR is always 0.5, which makes legitimate user-generated keys unpredictable by Eve.
In a wave-blockage environment, the issue of RISL attack becomes similar to the problem of untrusted relay, which could be solved by protocols based on private randomness \cite{untrusted}. 
Another possible way to mitigate the impact of RISL attack is to increase the number of helper RIS devices in the system. When multiple helper RIS are deployed, the number of multipaths between Alice and Bob will increase, resulting in a decrease in the proportion of leaked key information.

\section{CONCLUSION AND FUTURE DIRECTIONS}
In this work, we have reviewed PKG in view of the recently emerging RIS. %In particular, we have shown that the RIS can have constructive as well as destructive aspects on the PKG process. The RIS may act as a helper to assist or may be utilized by a malicious party to jam or (partially) control PKG. 
Based on our observations, we outline future directions.

\begin{itemize}
    \item \textit{The optimization of surface configurations in a secure and practical manner:} The current RIS literature pursues feedback of channel estimations. However, this is perpendicular to treating the channel response as a shared secret. Therefore, more work is required on how to optimally configure RIS for PKG without the optimization itself giving away useful information to eavesdroppers. Another open question remains on how to find strongly quantized RIS configurations in real-time to accomplish the remaining goals of SNR improvement and leakage reduction. %Thus, more work is required on practical RIS configuration techniques. Further, we assumed Alice and Bob adjust their channel probing to the RIS timing. Likewise, a protocol-aware RIS could adjust its timing to the user's channel probing.
    \item  \textit{The effects of RIS on other key generation processes and in frequency division duplex (FDD) systems:} The available literature mainly studies the role of RIS in the channel sounding stage of PKG. However, the influence of RIS on other key generation processes needs to be explored. For example, the RIS could help reduce information leakage during information reconciliation and quantization schemes may be optimized towards RIS-induced channel variations. Also, RIS-aided PKG in FDD systems is one of the promising topics in the future.
    \item  \textit{The optimal deployment of multiple RISs:} Compared with a single RIS, multiple RISs can greatly improve the time-variability of the channel. However, there is no literature considering the optimal deployment of RIS in PKG. Also, when a large number of RIS are controlled by the BS, jointly optimizing the reflection coefficients of multiple RISs at the same time would trigger an explosion of computational overhead, resulting in an intolerable key generation delay.
    \item  \textit{The detection and defense of RIS-based attacks:} Future work should investigate a general scenario where a malicious RIS and a friendly RIS exist concurrently. How can nodes distinguish between a friendly or a malicious RIS? It appears imperative to link the PKG process more to classical cryptography, allowing to authenticate nodes. However, it is yet to be explored, how this may be expanded to the physical layer to authenticate RIS signals.
\end{itemize}

\section{Acknowledgment}
This work was supported in part by the National Natural Science Foundation of China under Grant 6217011510, and in part by the German Research Foundation (DFG) within the framework of the Excellence Strategy - EXC 2092 CASA - 390781972, and in part by German Federal Ministry of Education and Research (BMBF) within the project MetaSEC (Grant 16KIS1234K). We thank our colleagues Markus Heinrichs and Prof. Rainer Kronberger at the High Frequency Laboratory, TH Cologne - University of Applied Sciences, Cologne, Germany, for providing the RIS prototype devices.

\bibliographystyle{IEEEtran}
\bibliography{IEEEabrv,ref}

\begin{thebibliography}{10}
\providecommand{\url}[1]{#1}
\csname url@samestyle\endcsname
\providecommand{\newblock}{\relax}
\providecommand{\bibinfo}[2]{#2}
\providecommand{\BIBentrySTDinterwordspacing}{\spaceskip=0pt\relax}
\providecommand{\BIBentryALTinterwordstretchfactor}{4}
\providecommand{\BIBentryALTinterwordspacing}{\spaceskip=\fontdimen2\font plus
\BIBentryALTinterwordstretchfactor\fontdimen3\font minus
  \fontdimen4\font\relax}
\providecommand{\BIBforeignlanguage}[2]{{%
\expandafter\ifx\csname l@#1\endcsname\relax
\typeout{** WARNING: IEEEtran.bst: No hyphenation pattern has been}%
\typeout{** loaded for the language `#1'. Using the pattern for}%
\typeout{** the default language instead.}%
\else
\language=\csname l@#1\endcsname
\fi
#2}}
\providecommand{\BIBdecl}{\relax}
\BIBdecl

\bibitem{2021Encrypting}
G.~Li, Z.~Zhang, J.~Zhang, and A.~Hu, ``Encrypting wireless communications on
  the fly using one-time pad and key generation,'' \emph{IEEE Internet Things
  J.}, vol.~8, no.~1, pp. 357--369, 2021.

\bibitem{AR}
G.~Li, A.~Hu, J.~Zhang, and B.~Xiao, ``Security analysis of a novel artificial
  randomness approach for fast key generation,'' in \emph{IEEE Glob. Commun.
  Conf.(GLOBECOM)}, Dec. 2017, pp. 1--6.

\bibitem{relay}
M.~Letafati, A.~Kuhestani, H.~Behroozi, and D.~W.~K. Ng, ``Jamming-resilient
  frequency hopping-aided secure communication for internet-of-things in the
  presence of an untrusted relay,'' \emph{IEEE Trans. Wireless Commun.},
  vol.~19, no.~10, pp. 6771--6785, 2020.

\bibitem{Low-entropy}
P.~Staat, H.~Elders-Boll, M.~Heinrichs, R.~Kronberger, C.~Zenger, and C.~Paar,
  ``{Intelligent Reflecting Surface-Assisted Wireless Key Generation for
  Low-Entropy Environments},'' in \emph{Proc. IEEE Int. Symp. Person. Indoor
  Mobile Radio Commun. (PIMRC)}, Virtual, Sep. 2021, pp. 1--7.

\bibitem{pan2021reconfigurable}
C.~Pan, H.~Ren, K.~Wang, J.~F. Kolb, M.~Elkashlan, M.~Chen \emph{et~al.},
  ``Reconfigurable intelligent surfaces for {6G} systems: Principles,
  applications, and research directions,'' \emph{IEEE Commun. Mag.}, vol.~59,
  no.~6, pp. 14--20, 2021.

\bibitem{Jirandom}
Z.~Ji, P.~L. Yeoh, G.~Chen, C.~Pan, Y.~Zhang, Z.~He \emph{et~al.}, ``Random
  shifting intelligent reflecting surface for {OTP} encrypted data
  transmission,'' \emph{IEEE Wireless Commun. Lett.}, vol.~10, no.~6, pp.
  1192--1196, 2021.

\bibitem{SPletter}
X.~{Lu}, J.~{Lei}, Y.~{Shi}, and W.~{Li}, ``Intelligent reflecting surface
  assisted secret key generation,'' \emph{IEEE Signal Process. Lett.}, pp.
  1--1, 2021.

\bibitem{Ji}
Z.~{Ji}, P.~L. {Yeoh}, D.~{Zhang}, G.~{Chen}, Y.~{Zhang}, Z.~{He}
  \emph{et~al.}, ``Secret key generation for intelligent reflecting surface
  assisted wireless communication networks,'' \emph{IEEE Trans. Veh. Technol.},
  pp. 1--1, 2020.

\bibitem{hu2021ris}
L.~{Hu}, G.~{Li}, H.~{Luo}, and A.~{Hu}, ``On the {RIS} manipulating attack and
  its countermeasures in physical-layer key generation,'' in \emph{Proc. IEEE
  Veh. Technol. Conf. (VTC fall)}, Virtual, Sep. 2021, pp. 1--5.

\bibitem{sum}
G.~Li, C.~Sun, W.~Xu, M.~D.Renzo, and A.~Hu, ``On maximizing the sum secret key
  rate for reconfigurable intelligent surface assisted multiuser systems,''
  \emph{accepted by IEEE Trans. Inf. Forensics Security}, 2021.

\bibitem{quantizer}
C.~Zenger, J.~Zimmer, and C.~Paar, ``Security analysis of quantization schemes
  for channel-based key extraction,'' \emph{EAI Endorsed Transactions on
  Security and Safety}, vol.~2, no.~6, p.~e5, 2015.

\bibitem{rukhin_statistical_2010}
A.~Rukhin, J.~Soto, J.~Nechvatal, M.~Smid, and E.~Barker, ``A {Statistical}
  {Test} {Suite} for the {Validation} of {Random} {Number} {Generators} and
  {Pseudo} {Random} {Number} {Generators} for {Cryptographic} {Applications},''
  NIST, {NIST} {Pubs} Special Publication (NIST SP) - 800-22 Rev 1a, 2010.

\bibitem{kraskovEstimatingMutualInformation2004}
A.~Kraskov, H.~Stögbauer, and P.~Grassberger, ``{Estimating Mutual
  Information},'' \emph{Physical Review E}, vol.~69, no.~6, p. 066138, 2004.

\bibitem{var}
N.~Patwari, J.~Croft, S.~Jana, and S.~K. Kasera, ``High-rate uncorrelated bit
  extraction for shared secret key generation from channel measurements,''
  \emph{IEEE Trans. Mob. Comput.}, vol.~9, no.~1, pp. 17--30, 2010.

\bibitem{untrusted}
N.~Aldaghri and H.~Mahdavifar, ``Physical layer secret key generation in static
  environments,'' \emph{IEEE Trans. Inf. Forensics Security}, vol.~15, pp.
  2692--2705, 2020.

\end{thebibliography}
\end{document}